\documentstyle[12pt]{article}
\newcommand{\be}{\begin{equation}}
\newcommand{\en}{\end{equation}}
\newcommand{\bea}{\begin{eqnarray}}
\newcommand{\ena}{\end{eqnarray}}
\newcommand{\Det}{\hbox{Det}}
\newcommand{\hbo}{\hbox to 1 true cm {\hfill } }
\newcommand{\tr}{\hbox{tr}}

\def\dslash{\partial\kern-.5em\slash}
\def\kslash{k\kern-.5em\slash}

\begin{document}
\vglue 1truecm

\vbox{
\hfill UNIT\"U-THEP-18/1992
}

\vfil
\centerline{\bf \large Instanton condensation in field strength
   formulated QCD$^1$}

\bigskip
\centerline{K.\ Langfeld, H.\ Reinhardt }
\medskip
\centerline{Institut f\"ur theoretische Physik, Universit\"at
   T\"ubingen}
\centerline{D--7400 T\"ubingen, F.R.G.}
\bigskip

\vfil
\begin{abstract}

Field strength formulated Yang-Mills theory is confronted with the
traditional formulation in terms of gauge fields.
It is shown that both formulations yield the same semiclassics,
in particular the same instanton physics.
However, at the tree level the field strength approach is superior
because it already includes a good deal of of quantum fluctuations
of the standard formulation.
These quantum fluctuations break the scale invariance of
classical QCD and give rise to an instanton interaction and this
causes the instantons to condense and form a
homogeneous instanton solid. Such the instanton solids show
up in the field strength approach as homogeneous (constant up to
gauge transformations) vacuum solutions. A new class of SU(N)
instantons is presented which are not embeddings of SU(N-1)
instantons but have non-trivial SU(N) color structure
and carry winding number $n=N(N^{2}-1)/6$. These instantons generate
(after condensation) the lowest action solutions of the
field strength approach. The statistical weight (entropy) of
different homogeneous solutions for SU(3) is numerically estimated by
Parisi's stochastic quantization method. Finally, we compare
instanton induced quark condensation with the condensation
of quarks in the homogeneous field strength solutions. Our investigations
show that the homogeneous vacuum of the field strength approach
simulates in an efficient way a condensate of instantons.

\end{abstract}

\vfil
\hrule width 5truecm
\vskip .2truecm
\begin{quote}
$^1$ Supported by DFG under contract Re $856/1 \, - \, 1$
\end{quote}
\eject
\tableofcontents
\section{Introduction}
\label{sec:1}
\medskip
Quantumchromodynamics (QCD) is widely accepted as the theory of
strong interactions. Our belief in QCD is mainly based on two facts:
first, it provides a successful description
at high energies where asymptotic freedom has been
experimentally verified. Secondly,
because it embodies the approximate chiral symmetry seen
in the low-energy hadron spectrum.

The low-energy sector of QCD is not fully understood yet, but
it is well established that the perturbative (continuum)
QCD vacuum becomes unstable at low energies~\cite{sav77}.
At low energies, the effective coupling constant is large, and
non-perturbative methods are required. The only rigorous approach
to the strong coupling regime of QCD are lattice simulations
based on Wilson's lattice formulation of Yang-Mills (YM) theory~\cite{wi74}.
However, due to renormalization, the lattice approach does
not coincide a priori with the continuum YM-theory.
It is only in the weak coupling limit that the renormalized lattice theory
exhibits the correct scaling behavior of perturbative YM-theory~\cite{cr80}.
Furthermore, the lattice approach requires very time consuming
numerical work yet the gained physical insight is rather meager.
Hence, alternative non-perturbative approaches to YM-theories
would be very welcome and in the past several proposes have been made:
for example the approach
of Pagels~\cite{pag}, the instanton gas approach of Callan, Dashen
and Cross~\cite{ca78}, the instanton liquid model
proposed by Shuryak~\cite{sh82}
and later elaborated in refs.\ \cite{dya83} and \cite{sh90}.

Instantons are generally believed to play an important role
in the QCD ground state and in particular, in the confinement
mechanism. Furthermore instantons (in both the dilute gas and instanton
liquid picture) may possibly trigger spontaneous breaking of chiral
symmetry and perhaps offer an explanation of the $U_{A}(1)$
problem~\cite{tho81}.

Recently, a somewhat different approach to low-energy QCD has been proposed
where the YM sector is entirely described in terms of the
field strength~\cite{ha77,sch90}, resulting in an effective tensor theory.
This so called field strength approach (FSA) offers a very
concise description of the QCD vacuum -
at the 'classical' (tree) level it is given by a homogeneous
condensate of gluons~\cite{sch90,al91}. When quantum fluctuations
are included~\cite{re90a}, such homogeneous condensates become unstable,
and tend to break into domains~\cite{am90}
of constant color electric and color magnetic field, reminiscent
of the Copenhagen vacuum~\cite{nie78}. Furthermore
the FSA results in an effective tensor theory, which already
exhibits an anomalous breaking of scale invariance at the tree level.
This is because the corresponding effective action of the
field strength contains an explicit energy scale, which means that
the action cannot tolerate the usual instantons as stationary points,
since instantons contain a free scale parameter.
One might wonder how the familiar instantons manage to escape in the FSA.

In this paper we confront the FSA with the more traditional instanton
picture of the QCD vacuum. We will find that the constant
chromo-electric and chromo-magnetic field configurations of the FSA represent
a coherent superposition of condensed instantons.
Individual instantons cease to exist as classical solutions
in the FSA since the effective action already
includes a good deal of the quantum fluctuations of the original
gluon field. It is these quantum fluctuations which trigger the condensation
of the instantons.

The remainder  of the paper is as follows:
In the next section we briefly review the field strength formulation
of YM-theory and compare the semiclassical description based on
instantons in the field strength formulation with that of the
standard formulation. Section
\ref{sec:3.1} presents a special class of SU(N) instantons which
are not embeddings of the SU(2) instanton
but have non-trivial SU(N) color structure.
In section \ref{sec:3} we confront the FSA with the instanton
picture. We find that in the field strength formulation
the SU(N) instantons condense and form homogeneous instanton
solids which appear as constant classical FSA solutions.
We then show that these instanton solids form the lowest action solutions
of the FSA in section \ref{sec:3.3}. We also give an
estimate of the statistical weight of classical FSA solutions
with higher actions by using Parisi's stochastic quantization method.
Section \ref{sec:4} compares  quark condensation in an
instanton background and in the homogeneous FSA solutions.
A short summary and some concluding remarks are given
in the final section.

\section{Semi-classical approximation to Yang-Mills theory}
\label{sec:2}
\medskip
In the following we concentrate on the gluon sector with the
quarks considered as spectators. The generating functional
for gluonic Green's functions of QCD is defined by
\footnote{ We discard explicitly gauge fixing, which will not
alter the following considerations. It can be implemented afterwards
by restricting the class of instantons.}:
\be
Z[j] \; = \; \int {\cal D} A \; \exp ( - \frac{ 1 }{ 4 g^{2} }
\int d^{4}x \; F^{a}_{\mu \nu }[A] F^{a}_{\mu \nu }[A] \, + \,
\int d^{4}x \; j^{a}_{\mu } A^{a}_{\mu } \, ) \; ,
\label{eq:1}
\en
where $j^{a}_{\mu }$ is the color octet quark current which
acts like an external source in the gluon sector, and
\be
F^{a}_{\mu \nu }[A] \; = \; \partial _{\mu } A^{a}_{\nu } \, - \,
\partial _{\nu } A^{a}_{\mu } \, + \, f^{abc} A^{b}_{\mu }
A^{c}_{\nu }
\label{eq:2}
\en
is the field strength of the gauge field $A^{a}_{\mu }$.
The integration over gauge field degrees of freedom
cannot be performed exactly and
one has to resort to approximations. At high energies,
a perturbative expansion in terms of the coupling strength is appropriate.
However, at low energies the coupling constant is expected to be large
and non-perturbative methods are required.
One possibility is to resort to a semi-classical description,
expanding (\ref{eq:1}) around the non-trivial gauge field configurations
which are known to exist and referred to as instantons.
This yields the instanton picture
extensively discussed in the literature~\cite{inst}.
We will return to this approach in section \ref{sec:2.1}.

A different approach to low energy QCD
results if one reformulates (\ref{eq:1}) in terms of
the field strength~\cite{ha77,sch90}. For this purpose
we introduce the
field strength as independent variable by inserting the identity
\be
1 = \int {\cal D} G \; \delta (G - F[A]) \; = \;
\int {\cal D} G \;
{\cal D} T \; \exp [ + \frac{ i }{ 2 } \int d^{4}x \; T^{a}_{\mu \nu }
( G^{a}_{\mu \nu }- F^{a}_{\mu \nu }[A]) ]
\label{eq:3a}
\en
into the generating functional $Z[j]$.  The path integral (\ref{eq:1})
then becomes
\bea
Z[j] &=& \int {\cal D} A \; {\cal D} G \; {\cal D} T \; \exp (
- \int d^{4}x \; L )
\label{eq:3} \\
L &=& \frac{1}{4 g^{2}} G^{a}_{\mu \nu } G^{a}_{\mu \nu }
- \frac{i}{2} G^{a}_{\mu \nu } T^{a}_{\mu \nu }
- \frac{i}{2} A^{b}_{\mu } \hat{T}^{bc}_{\mu \nu } A^{c}_{\nu }
- (i \partial _{\mu } T^{a}_{\mu \nu } + j^{a}_{\nu }) A^{a}_{\nu } \; ,
\label{eq:4}
\ena
where $\hat{T}^{ab}_{\mu \nu } = f^{abc} T^{c}_{\mu \nu }$. In the
following we will assume $\det \, \hat{T} \not= 0 $ since
configurations with $\det \, \hat{T} = 0$ are statistically
suppressed. Following~\cite{sch90}, integration over
the gauge field $A^{a}_{\mu }$ and field strength $G^{a}_{\mu \nu }$ yields
\bea
Z[j] &=& \int {\cal D} T \; {\cal G}[j] \; \Det ^{-1/2} \hat{T} \;
\exp [ - \int d^{4}x \; L_{T} ]
\label{eq:12} \\
L_{T} &=& \frac{g^{2}}{4} T^{a}_{\mu \nu } T^{a}_{\mu \nu } -
\frac{i}{2} \partial _{\omega } T^{a}_{\omega \mu }
(\hat{T}^{-1})^{ab}_{\mu \nu } \partial_{\sigma } T^{b}_{\sigma \nu } \; ,
\nonumber \\
{\cal G}[j] &=& \exp [ j^{a}_{\mu } (\hat{T}^{-1})^{ab}_{\mu \nu }
\partial _{\omega } T^{b}_{\omega \nu } \, - \, \frac{i}{2}
j^{a}_{\mu } (\hat{T}^{-1})^{ab}_{\mu \nu } j^{b}_{\nu } ] \; .
\label{eq:13}
\ena
Equations (\ref{eq:12},
\ref{eq:13}) are an exact reformulation of the gluonic generating functional
(\ref{eq:1}) as a path integral over the conjugate field $T^{a}_{\mu \nu }$.
Note that the integral over $T^{a}_{\mu \nu }$
in (\ref{eq:12}) can no longer be evaluated in closed form. However,
recent investigations show that a semiclassical treatment is
appropriate (see e.\ g.\ \cite{re91,la91a,la91b,al92}).

\bigskip
\subsection{Instanton picture}
\label{sec:2.1}
\medskip

In the standard continuum approach to the YM ground state
a semiclassical analysis of the path integral over the gauge field
$A_{\mu }$ is performed. Below
we demonstrate that the same semiclassical approximation is
obtained in the field strength formulation, as was recently
shown~\cite{re92}. Extremizing the YM action
\be
S_{YM} \; = \; \frac{1}{ 4 g^{2} } \int d^{4}x \;
F^{a}_{\mu \nu }[A] F^{a}_{\mu \nu }[A]
\label{eq:yma}
\en
yields
the classical equation of motion:
\be
\partial _{\nu } F^{a}_{\nu \mu }[A] \; + \; f^{abc} A^{b}_{\nu }
F^{c}_{\nu \mu }[A] \; = \; 0 \; .
\label{eq:15a}
\en
Non-trivial finite action solutions to this equation are
usually referred as instantons.
These instanton configurations also show up in the field strength
formulation~\cite{ha77,re92}.
Variation of the exponent in (\ref{eq:12}) with
respect to the conjugate field $T^{a}_{\mu \nu }$ yields the classical
equation of motion
\be
g^{2} \, T^{a}_{\mu \nu } \; + \; i F^{a}_{\mu \nu }[V] (x)
\; = \; 0 \; ,
\label{eq:11}
\en
where $F^{a}_{\mu \nu }[V]$ denotes the field strength (\ref{eq:2})
of the vector field
\be
V^{a}_{\mu }[T] \; = \; (\hat{T}^{-1})^{ab}_{\mu \nu }
\partial _{\omega } T^{b}_{\omega \nu } \; .
\label{eq:10}
\en
This quantity transforms under changes of gauge in the same manner
as the original gauge potential $A_{\mu }$ and has only a linear
coupling to the quark current, also in common with $A_{\mu }$.
It should therefore be understood as the counterpart of the gauge potential
in the field strength formulation.
Equation (\ref{eq:11}) is solved by the field strength
of an instanton $A^{\mu }_{inst}$~\cite{re92}
\be
T^{a}_{\mu \nu } = - \frac{i}{ g^{2} } F^{a}_{\mu \nu } [A_{inst}]
\label{eq:11x}
\en
with the corresponding induced vector field
(\ref{eq:10})
\be
V^{a}_{\mu }[T] \, = \, V^{a}_{\mu }[ \, - \frac{i}{g^{2}} F^{a}_{\mu \nu }
[A_{inst}] \, ] = \, V^{a}_{\mu }[  F^{a}_{\mu \nu } [A_{inst}] \, ]
\, = \, A_{\mu }(x)_{inst}
\label{eq:11y}
\en
which coincides with the instanton potential.

The classical solutions contribute to the generating functional
$Z[j]$ with a weight factor given by the
exponent of the classical instanton action times the integral over
small fluctuations. The classical instanton action is the same
in both approaches (\ref{eq:1}) and (\ref{eq:12})~\cite{re92}.
In fact, the tensor action in (\ref{eq:12}) can be written as
\be
\int d^{4}x \; L_{T} \; = \; \int d^{4}x \; \{ \frac{ g^{2} }{4} T^{2}
\, + \, \frac{i}{2} T^{a}_{\mu \nu } F^{a}_{\mu \nu } [V] \; \} \; ,
\label{eq:11c}
\en
where we have used
\be
\int d^{4}x \;
\partial _{\omega } T^{a}_{\omega \mu }
(\hat{T}^{-1})^{ab}_{\mu \nu } \partial_{\sigma } T^{b}_{\sigma \nu } \; = \;
- \int d^{4}x \; T^{a}_{\mu \nu } F^{a}_{\mu \nu }[V] \; ,
\label{eq:2.1.1a}
\en
Inserting  (\ref{eq:11x}) into (\ref{eq:2.1.1a}) gives
the field strength action of an instanton i.\ e.\ ,
\be
\int d^{4}x \; L_{T} \; = \; \int d^{4}x \;  \frac{ 1 }{4 g^{2} }
F^{a}_{\mu \nu } [V] F^{a}_{\mu \nu } [V] \; ,
\label{eq:11d}
\en
which agrees with the standard YM action (\ref{eq:yma}).

It remains to compare the weight factors obtained from
the integration over the modes of the fluctuations. In the standard
formulation this gives rise to a weight factor
\be
Q ( \Det ' \frac{ \delta ^{2} S }{ \delta A^{a}_{\mu } \delta A^{b}_{\nu } }
) ^{-1/2} \; ,
\label{eq:11a}
\en
where the prime indicates that the zero modes are to be excluded from
the determinant. The zero mode contribution is denoted by $Q$.
In the field strength formulation the weight factor
is given by
\be
Q_{FS} ( \Det i \hat{T} )^{-1/2} \; (\Det ' \frac{ \delta ^{2} S_{T} }{
\delta \hat{T}^{a}_{\mu \nu } \delta  \hat{T}^{b}_{\mu \nu } }
)^{-1/2} \; .
\label{eq:11b}
\en
The second factor is here already present at the tree level whereas the
first and the last factor arise from the integration of fluctuations around
classical instanton field strength configurations.
It has been shown recently~\cite{re92} that  both determinants
(\ref{eq:11a}) and (\ref{eq:11b})
are equal (up to an irrelevant constant).
Hence the field strength formulation
reproduces the correct semiclassical description of the standard formulation
of YM theory.
The approaches differ, however, at tree level,
where the field strength formulation yields a pre-exponential
weight factor, $\Det ^{-1/2} \hat{T}$,
and an additional quark current-current
interaction, see (\ref{eq:12}), (\ref{eq:13}).
This weight factor arises from the integration
over the gauge fields in the field strength formulation and
so already includes some of the effects of the fluctuations
of the gauge field around an instanton configuration.
It should also be noted that the
additional quark interaction in (\ref{eq:12}) and (\ref{eq:13})
has a similar structure to the effective quark interaction which arises
in a correlated instanton anti-instanton gas~\cite{ca78,sh82}.
One therefore expects the field strength formulation to already
describe the dominant quantum and correlation effects of the instantons
at tree level. We will provide further evidence for this later.
This fact makes the field strength formulation very attractive
for studying the low energy sector of YM theories.

\bigskip
\subsection{ Effective tensor theory (FSA) }
\label{sec:2.2}
\medskip
The field strength formulation of non-abelian YM-theories
(\ref{eq:12},\ref{eq:13}) offers a concise non-perturbative
treatment if one includes the weight factor, $\Det ^{-1/2} \hat{T}$,
into the action.
This yields the so-called field strength approach (FSA) to
YM-theories~\cite{sch90}.
After renormalization, the generating functional
of the field strength approach takes the form~\cite{sch90}
\be
Z[j] \; = \; \int {\cal D}T \; {\cal G} [j] \;
\exp [ - \int d^{4}x L_{FSA} ] \; ,
\label{eq:50}
\en
where the effective Lagrangian is given by
\be
L_{FSA} \; = \; \frac{ g^{2} }{4} T^{2} - \frac{i}{2}
\partial _{\omega } T^{a}_{\omega \mu } ( \hat{T} ^{-1} )^{ab}_{\mu \nu }
\partial _{\sigma } T^{a}_{\sigma \nu }
+ \frac{ \mu ^{4} }{2} \tr \ln \hat{T}(x)
\label{eq:51}
\en
and the source term ${\cal G}[j]$ (\ref{eq:13})
remains unchanged.
The renormalization procedure induces the energy scale $\mu $
at tree level reflecting the presence of the scale anomaly~\cite{co77}.
The physical value of $\mu $ has to be fixed by a renormalization
condition. For instance, in the classical approximation, $\mu $ is
related to the non-perturbative gluon condensate~\cite{sch90}.
The classical equation of motion of the
tensor field then becomes (cf.\ eq.\ (\ref{eq:11}))
\be
g^{2} T^{a}_{\mu \nu } \; + \; \mu ^{4}
f^{abc} ( \hat{T}^{-1} )^{ab}_{\mu \nu } \; + \; i \,
F^{a}_{\mu \nu } [V] \; = \; 0 \; ,
\label{eq:52}
\en
Due to the presence of the explicit energy scale $\mu $ one might expect that
instantons with arbitrary size no longer exist.
In fact, the effective action (\ref{eq:51}) no longer has instanton
solutions at all. Instead the non-trivial stationary points
are given in a certain gauge by space-time constant tensor field
configurations~\cite{sch90}. These homogeneous
classical solutions appear at the expense of the instantons.
One is tempted to interpret these constant solutions as
a conglomerate of condensed instantons.
In the following sections we give further support to this
interpretation. In fact we demonstrate that the constant
SU(N) field strength solutions of minimal classical action
have the same color and Lorentz structure as certain SU(N)
instantons. First of all  we describe these SU(N)
instanton solutions.

\bigskip
\section{ SU(N) instanton solutions }
\label{sec:3.1}
\medskip
There is a general method based on works of Atiyah et
al.\ \cite{at77,co78} which, in principle, allows one to find all (anti-)
self-dual instanton solutions, although in practice it might be
difficult to obtain explicit representations. In order to compare
the FSA with the more traditional instanton physics we will be
interested in a certain class of instantons for which we
need explicit representations satisfying simple algebraic
relations. For this purpose it is more convenient to follow
a different route which immediately provides the explicit  representation
of the instantons. Actually, we only became aware of the work of
Atiyah et al.\
after we had already found the instantons relevant in the
context of the FSA.

\bigskip
\subsection{ Instantons in Schwinger-Fock gauge}
\label{sec:a}
\medskip
Instantons are finite action extrema of the classical Euclidean
Yang-Mills action (\ref{eq:1}) which solve
the classical equations of motion:
\be
\partial _{\mu } F^{a}_{\mu \nu } \; = \; \hat{F}^{ab}_{\nu \mu }
A^{b}_{\mu } \; , \hbo
\hat{F}^{ab}_{\mu \nu } \; = \; f^{abc} F^{c}_{\mu \nu } \; .
\label{eq:1a}
\en
Any finite action gauge potential $A^{a}_{\mu }(x)$, in particular
the instantons, must become pure gauge $U \partial _{\mu } U^{\dagger }$
as $x^{2}$ tends to infinity. They can be classified
by the so called topological charge
\be
n [F] \; = \; \frac{ 1 }{ 16 \pi ^{2}  } \int d^{4}x \;
\tr ( F_{\mu \nu } F^{\ast}_{\nu \mu } )
\label{eq:6i}
\en
which is  an integer (the Pontryagin index).
The Bianchi-identity
\be
\partial _{\mu } F^{a \, \ast }_{\mu \nu } \; = \; (\hat{F} ^{\ast }
 )^{ab}_{\mu \nu } A^{b}_{\mu } \; , \hbo
F^{a \, \ast }_{\mu \nu } \; = \; \frac{1}{2} \epsilon _{\mu \nu
\alpha \beta } F^{a  }_{\alpha \beta } \; ,
\label{eq:2a}
\en
which holds for arbitrary gauge fields, implies that any gauge field
which has an (anti-) self dual field strength (i.\ e.\
$F^{a \, \ast }_{\alpha \beta } = (-) F^{a }_{\alpha \beta }$ ),
represents an instanton solution to the classical YM equation (\ref{eq:1a}),
in agreement with the standard variational result~\cite{inst}.

For a given gauge potential $A^{a}_{\mu } (x)$ however, we do not know
whether its field strength $F^{a}_{\mu \nu }$ is self-dual
until we actually calculate it. Thus in the search for instantons,
it is much more convenient to consider the field strength
as an independent dynamical variable instead of the gauge potential
itself. This can be accomplished in the Fock-Schwinger gauge
\be
x_{\mu }  A_{\mu }(x)=0 \; .
\label{gauge}
\en
In this gauge,
the gauge potential can be entirely expressed in terms of the
field strength i.\ e.\ ,
\be
A^{a}_{\mu }(x) \; = \; - \int_{0}^{1} d\alpha \;
\alpha \; F^{a}_{\mu \nu }(\alpha x ) \; x_{\nu }  \; .
\label{eq:a1}
\en
and the equation of motion (\ref{eq:1a}) becomes an
equation for $F^{a}_{\mu \nu }$ i.\ e.\ ,
\be
\partial _{\mu } F^{a}_{\mu \nu }(x) \; = \; - \hat{F}^{ab}_{\nu \lambda }
(x) \int_{0}^{1} d \alpha \; \alpha \, F^{b}_{\lambda \sigma }
x_{\sigma } \; .
\label{eq:a1a}
\en
This equation is equivalent to the original equation of motion only
for those $F^{a}_{\mu \nu } $ which are the field strengths to some
gauge potential. For an arbitrary $F'^{a}_{\mu \nu }$
equation (\ref{eq:a1}) yields a gauge potential $A^{a}_{\mu }[F']$
whose field strength $F^{a}_{\mu \nu }\left[ A[F'] \right] $
differs in general from the initial
$F'^{a}_{\mu \nu }$. We must therefore supplement (\ref{eq:a1a})
with the constraint
\be
F^{a}_{\mu \nu } \bigl[ A[F] \bigr] \; = \; F^{a}_{\mu \nu } \; .
\label{eq:constr}
\en
With this constraint, (\ref{eq:a1a})
is completely equivalent to the classical YM equation
of motion (\ref{eq:1a}) that we started with.

We now solve (\ref{eq:a1a}) with the isotropic ansatz
\be
F^{a}_{\mu \nu } \; = \; G^{a}_{\mu \nu } \, \psi (x^{2}) \; ,
\label{eq:a2}
\en
where $G^{a}_{\mu \nu } $ is a constant, (anti-) self-dual matrix
antisymmetric under the exchange of $\mu $ and $\nu $.
For this ansatz the gauge field (\ref{eq:a1}) becomes
\be
A^{a}_{\mu } \; = \; - G^{a}_{\mu \nu } x_{\nu } \frac{1}{2 x^{2} }
\int_{0}^{x^{2} } du \; \psi (u) \; =: \;
- G^{a}_{\mu \nu } x_{\nu } \Phi ( x^{2} ) \; .
\label{eq:a3}
\en
and the equation of motion (\ref{eq:a1a}) reduces to
\be
- 2 G^{a}_{\mu \nu } x_{\nu } \, \psi ' \; + \; \hat{G}^{ab}_{\mu \nu }
G^{b}_{\nu \omega } x_{\omega } \, \Phi (x^{2}) \psi (x^{2}) \; = \; 0 \; .
\label{eq:a5}
\en
This equation makes it easy to separate
the color and Lorentz structure from the space-time
dependence, reducing (\ref{eq:a5})
(uniquely up to unimportant rescaling of $G \rightarrow G/ \beta $
and $\Phi \rightarrow \beta \Phi $ with an arbitrary constant $\beta $) to
\bea
\hat{G}^{ab}_{\mu \lambda } G^{b}_{\lambda \nu } &=& 2 \, G^{a}_{\mu \nu }
\label{eq:a6} \\
\psi '( x^{2} ) & = & \Phi (x^{2}) \, \psi (x^{2}) \; .
\label{eq:a7}
\ena
%
In order to solve the matrix equation (\ref{eq:a6}) it is convenient to expand
the antisymmetric color matrices $G^{a}_{\mu \nu }$ in terms
of 't~Hooft's $\eta $-symbols i.\ e.\ ,
\be
G^{a}_{\mu \nu } \; = \; G^{a}_{i} \eta^{i}_{\mu \nu } \; + \;
\bar{G} ^{a}_{i} \bar{\eta }^{i}_{\mu \nu } \; .
\label{eq:14i}
\en
The $\eta ^{i} $ and $\bar{\eta }^{i}, \; (i=1,2,3)$ are self-dual
and anti self-dual space-time tensors, which generate the
$SU(2) \times SU(2) \sim SO(4)$ symmetry group of
Euclidean space, satisfying
\be
[ \eta ^{i}, \eta ^{k} ] = -2 \epsilon ^{ikl} \eta ^{l} \; ,
\hbox to 1 true cm {\hfill}
[ \bar{\eta } ^{i}, \bar{\eta } ^{k} ] = -2 \epsilon ^{ikl}
\bar{\eta } ^{l} \; ,
\hbox to 1 true cm {\hfill}
[ \eta ^{i}, \bar{\eta }^{k} ]=0  \; ,
\label{eq:15i}
\en
\be
\{ \eta ^{i}, \eta ^{k} \} \; = \; - 2 \delta ^{ik} \; , \hbo
\{ \bar{\eta }^{i}, \bar{\eta }^{k} \} \; = \; - 2 \delta ^{ik} \; , \hbo
\tr \{ \eta^{i} \bar{\eta }^{k} \} \; = \; 0 \; .
\nonumber
\en
For the moment let us confine ourselves to self-dual configurations
i.\ e.\ , $\bar{G}^{a}_{i}=0$.
Using (\ref{eq:15i}) the relation (\ref{eq:a6}) can be cast in the form
\be
G^{a}_{l} = \frac{1}{2} \epsilon _{lrs} f^{abc} G^{b}_{r} G^{c}_{s} \; .
\label{eq:32}
\en

We must also guarantee that the field strength ansatz (\ref{eq:a2})
is constructed from the gauge potential (\ref{eq:a3}).
So, our task is to solve the constraint (\ref{eq:constr}) together with
the e.o.m.'s (\ref{eq:a7}) and (\ref{eq:32}) for $F^{a}_{\mu \nu }$
and $A^{a}_{\mu }$ given by (\ref{eq:a2}) and (\ref{eq:a3})
respectively.
Decomposing the constraint (\ref{eq:constr}) into self-dual and
anti self-dual parts one finds
\bea
G^{a}_{i} ( 2 \Phi + x^{2} \Phi ' ) \, + \, \frac{1}{4}
\epsilon _{ikl} f^{abc} G^{b}_{k} G^{c}_{l} \, x^{2} \Phi ^{2}  &=&
G^{a}_{i} \psi
\nonumber \\
- x_{\nu } \bar{\eta } ^{i}_{\nu \mu } \eta ^{k}_{\mu \omega }
x_{\omega } \, ( G^{a}_{k} \Phi ' - \frac{1}{4}
\epsilon _{kml} f^{abc} G^{b}_{m} G^{c}_{l} \, \Phi ^{2} ) &=& 0 \; .
\nonumber
\ena
Using the relation (\ref{eq:32}) of the coefficients $G^{a}_{i}$, we
can reduce these to ordinary differential equations i.\ e.\ ,
\bea
x^{2} \Phi ' \, + 2 \Phi \, + \, \frac{1}{2} x^{2} \Phi ^{2}
&=& \psi
\label{eq:a15} \\
\ \Phi ' \, - \, \frac{1}{2} \Phi ^{2} &=& 0 \; .
\label{eq:a16}
\ena
These equations are solved by
\be
\Phi (x^{2}) \; = \; - \frac{2}{ x^{2} + \rho ^{2} } \; , \hbo
\psi (x^{2}) \; = \; - \frac{ 4 \rho ^{2} }{ ( x^{2} + \rho ^{2} )^{2} }
\label{eq:a17}
\en
where $\rho $ is an arbitrary parameter.
It is not surprising that these solutions also satisfy (\ref{eq:a7}),
which follows from the classical YM equation (\ref{eq:1a})
within the gauge (\ref{gauge}) and with the ansatz (\ref{eq:a2}).
This is merely a consequence of the general fact that
any (anti-) self-dual field configuration satisfies the classical
YM equations. In fact, the solutions (\ref{eq:a17}) have precisely the
space-time dependence of the gauge potential and the field strength
of the standard Polyakov-t'Hooft instanton.

What remains to be done is to solve the algebraic equation
(\ref{eq:32}) which defines the color and Lorentz structure
of the instantons.
Let $t^{a}$ denote the generators
of the gauge group $\cal G$ satisfying $[t^{a},t^{b}]=if^{abc} t^{c}$ in
the fundamental representation. Defining
\be
G_i \; = \; G^{a}_{i} t^{a} \; ,
\hbox to 1 true cm {\hfill}
\bar{G}_{i} \; = \; \bar{G}^{a}_{i} t^{a}
\label{eq:18i}
\en
the algebraic equation (\ref{eq:32})  can be rewritten as
\be
[ G_{j}, G_{k} ]  \; = \; i \epsilon _{jkl} G_{l} \; ,
\label{eq:34}
\en
Since the $\epsilon ^{ijk}$ are the structure constants of the SU(2)
group, any realization $G_{i}$
of the SU(2) algebra yields an instanton solution.

For field configurations (\ref{eq:a2}) with radial dependence
(\ref{eq:a17}) the winding number and the corresponding action become
\be
n \; = \; \frac{1}{3} \,  G^{a}_{i} G^{a}_{i} \; = \;
\frac{2}{3} \tr G_{i} G_{i} \; ,
\hbox to 1 true cm {\hfill}
S \; = \; \frac{ 8 \pi ^{2} }{ 3 g^{2} } \,  G^{a}_{i} G^{a}_{i} \; = \;
\frac{4 \pi ^{2} }{g^{2}} n \; .
\label{eq:20i}
\en
Noting that instanton solutions $G_{i}$
are representations of the SU(2) spin algebra for some spin $s$,
the quantity
$G_{i} G_{i}$ represents the quadratic
casimir operator of the SU(2) group, i.\ e.\ ,
\be
G_{i} G_{i} \; = \;  s(s+1) \; 1_{G} \; =: \; c(G) \, 1_{G}
\hbox to 2 true cm {\hfil with \hfil }
\tr \, 1_{G} =: d(G) \; .
\label{eq:21i}
\en
$1_{G}$ is the unit matrix in the spin representation defined
by the $G_{i}$,
which need not to coincide with the $N$-dimensional unit matrix of the color
representation but may have lower dimension.
The winding number and action can therefore be expressed as
\be
n= \frac{2}{3}  c(G) d(G) \; ,
\hbox to 1 true cm {\hfill}
S=\frac{16 \pi^{2} }{3 g^{2} }  c(G) d(G)  \; .
\label{eq:23i}
\en

\bigskip
\subsection{ Explicit construction of SU(N) instantons }
\label{sec:3.1.2}
\medskip
Let us now construct explicit solutions to (\ref{eq:34}), which
define the color structure of the instantons.
In the fundamental representation of SU($N$) the $t^{a}$
form a complete basis for the hermitean $N \times N$ matrices.
Furthermore, the lowest dimensional irreducible spin $s$ representation
of SU(2) is realized by hermitean $(2s+1) \times (2s+1)$
matrices. Hence, the N-dimensional $s=(N-1)/2$ spin representation always
provides an instanton $G_{i}$ with non-trivial SU($N$) color structure.
We refer to this instanton as the instanton of maximum spin (for given $N$).
{}From (\ref{eq:23i}) it follows that this
instanton carries winding number
\be
n \; = \; \frac{2}{3} \, N \, s(s+1) \; = \;
\frac{2}{3} N \, \frac{N-1}{2} \, \frac{N+1}{2} \; = \;
\frac{1}{6} N(N^{2}-1) \; .
\label{eq:29i}
\en
Additional instantons will usually arise from $N$-dimensional
realizations of lower spin $s<(N-1)/2$ representations. In particular,
all embeddings of SU($N-1$) instantons in SU($N$) obviously represent
SU($N$) instanton configurations.

For the gauge group SU(2) any representation $t^{a}$ of the color
generators naturally provides an instanton by identifying
the color group with the spin group, i.\ e.\ ,
\be
G^{a}_{i} \; = \; \delta ^{a}_{i} \hbo
(\hbox{ identity map }) \; .
\label{eq:24i}
\en
Eq.(\ref{eq:a2},\ref{eq:14i}) then yields
the standard Polyakov-'t~Hooft instanton
$(G^{a}_{\mu \nu }= \eta ^{a}_{\mu \nu })$.
This is the instanton with maximal spin for SU(2).
Obviously, in this case, there are
(up to global color and Lorentz transformations)
no mappings of the SU(2) color
group into the SU(2) spin algebra other than the identity map and
therefore no other instantons of the type (\ref{eq:a2}) exist.

In the case of a SU(3) gauge group the $G_{i}$ (\ref{eq:18i})
are hermitian $3 \times 3$
matrices. The SU(3) embedding of the SU(2) instanton forms a
SU(3) instanton configuration
\be
G^{a}_{i}= \delta ^{a}_{i} \; , \hbo a=1,2,3 \hbo
G^{b}_{i}=0 \; , \hbo b=4,\ldots , 8 \; .
\label{eq:25i}
\en
This is the known SU(3) instanton and is the basis of
existing  instanton models of the QCD vacuum~\cite{inst}.
This instanton corresponds to the $s=1/2$ representation of the
spin group and is illustrated in figure (5).
There also exists another non-trivial SU(3) instanton,
corresponding to the $s=1$ representation of the spin group, i.\ e.\ ,
\be
(\, G_{i} \, )_{kl} \; = \; -i \, \epsilon _{ikl}
\label{eq:26i}
\en
which is the maximal spin representation in $N=3$ dimensions.
These $G_{i}$ are up to a phase the antisymmetric generators
$t^{a}=2,5,7$ of the fundamental SU(3) representation, which also
forms the $\bar{3}$ representation of SU(3). For this  SU(3)
solution, the non-zero components read (see figure (5))
\be
G^{7}_{1} \; = \; - \, G^{5}_{2} \; = \; G^{2}_{3} \; = \; 2 \; .
\label{eq:27i}
\en
This solution thus has a non-trivial SU(3) color structure and
from (\ref{eq:23i}) it carries winding number $n=4$.
Since in three dimensions (besides the trivial $s=0$) only the
$s=1/2$ and $s=1$ representations of SU(2) can be realized there are no
further self-dual SU(3) instantons of type (\ref{eq:a2}).

For a SU(4) gauge group the embeddings of the SU(3) instantons
discussed above are trivially SU(4) instanton solutions. A
SU(4) instanton with non-trivial SU(4) color structure again arises
from the identification of the $G_{i}$'s with the maximal spin
$s=3/2$ representation of the SU(2) spin group in $N=4$ dimensions.
According to (\ref{eq:29i}) this solution
carries winding number $n=10$. Two further SU(4) solutions are provided
by the two $s=1/2$ representations of SU(2) in four dimensions
given by  the  't~Hooft's symbols
\be
G_{k} \; = \; - \frac{i}{2} \eta ^{k}  \hbox to 2cm {\hfill or \hfill }
G_{k} \; = \; - \frac{i}{2} \bar{\eta }^{k}
\label{eq:28i}
\en
when the $\eta^{k}_{\mu \nu }$, $\bar{\eta }^{k}_{\mu \nu }$
are now matrices in color space.
These instantons correspond to the $(\frac{1}{2}, 0)$ and
$(0, \frac{1}{2})$ representations of the SO(4)$\sim $SU(2)$\times $SU(2)
subgroup of SU(4).
{}From (\ref{eq:21i}) and (\ref{eq:23i}) we see that these instantons carry
winding number $n=2$.

Of course, to any self-dual instanton solution with
winding number $n$ found above there exists an anti self-dual
instanton with winding number $-n$ constructed by exchanging
$G_{i}$ and $\bar{G}_{i}$.

\bigskip
\section{ Confrontation of the FSA with the instanton picture }
\label{sec:3}
\medskip
In this section we demonstrate, using the special class
of instantons constructed in the previous section,
that the homogeneous field strength vacuum solutions of the FSA
represent solids of condensed instantons.
First, we show that two instantons with equal orientation
posses an attractive interaction. Second, we proof
that only constant solutions of the classical equation of motion exist
in the FSA.
We then argue
that superpositions of instantons with uniform orientation,
averaged over positions and sizes, form a homogeneous, constant
field strength vacuum solution.

\subsection{ Estimate of the instanton interaction }
\label{sec:3.2a}
\medskip
In order to investigate the effect of quantum fluctuations
on the instantons  we need to examine
the effect of the pre-exponential factor
$\Det ^{-1/2} \hat{T}$ in (\ref{eq:12}) and (\ref{eq:13}),
treating it
as a weight factor for instanton configurations. This determinant is
divergent and needs regularization and renormalization.
We adopt a simplified renormalization procedure
motivated by physical arguments, and which
turns out to be sufficient at our level of approximation.
In particular, for instanton configurations
$T^{a}_{\mu \nu } = -iF^{a}_{\mu \nu } $, the renormalized determinant
yields the same $\rho $-dependence as the standard
fluctuation determinant calculated by t'Hooft~\cite{tho76} (see also
\cite{ber79}). This $\rho $-dependence
implies that instantons with large radii $\rho $ compared
to the fundamental (renormalization) scale $1/\mu $
are preferred. This fact does not change if one considers two
instantons and their binary interaction as we shall see.

We now investigate the interaction of two separated instantons
within the FSA by comparing their tree level effective
action to that of two isolated instantons.
We confine ourselves to the simplest case of two
maximum spin instantons
with the same radius and orientation. We find that for
medium radii the potential contains
short-range repulsive and medium-range attractive parts, while
for large instanton radii the potential becomes purely attractive.
Since instantons with large radii are preferred,
instantons will fuse and form a solid.

As discussed in section \ref{sec:2.1}, once the
fluctuations of the $T$-field are included, the full instanton determinant
of the standard formulation is recovered.
In this case there is no need to regularize $\Det^{-1/2} \hat{T} $
separately. However, once this term is renormalized, it already contains
quantum effects which are physically relevant at low energies.
In particular, it already describes the scale anomaly
as we shall demonstrate shortly.

We start from the generating functional for Green's functions
(\ref{eq:12}), where we now include the weight factor $\Det ^{-1/2} \hat{T}$
in the action. Adopting lattice regularization we obtain
\be
\Det ^{-1/2} \hat{T} \; = \; \exp \{ - \frac{1}{2} \sum_{i} \tr \ln
\hat{T}(x_{i}) \} \; = \; \exp \{ - \frac{1}{2} \int d^{4}x \;
\mu _{0}^{4} (x) \, \tr \ln \hat{T}(x)/\mu ^{2} \} \; ,
\label{eq:2.1.0}
\en
where $\mu $ is a reference scale and
$\mu_{0}(x)$ is a divergent measure function reflecting the fact
that the sum over space-time points has been replaced by a
Riemann integral. For equally spaced lattice points
the regulator is $\mu _{0}(x) = 1/a^{4}$
with $a$ the lattice spacing. The ambiguity of the space-time dependence
of the regulator function $\mu _{0}(x) $ is due to the fact
that one is dealing with the determinant of a purely local operator.
However, such ambiguity can be removed by including fluctuations
in the field strengths which induce non-local terms.
At our level of approximation
it is sufficient to remove this ambiguity on physical grounds.
We demand that first the correct anomaly is generated by scale
transformations, and second that the regularization preserves the
individual particle picture of instantons. As an initial estimate
we follow the renormalization adopted in the
FSA~\cite{sch90}. One finds that $\mu _{0}$ is related, in the
classical approximation, to the gluon condensate in the vacuum.
As our renormalization condition, we retain
the relation between $\mu_{0}$ and the field strength
for space-time dependent instanton configurations and absorb
the divergence of $\mu _{0}$ by a rescaling of the
field strength.
We therefore generalize to space-time dependent
field configurations by setting the regulator function as
\be
\mu _{0}^{4 }(x) \; = \; \omega F^{a}_{\mu \nu }(x) F^{a}_{\mu \nu }(x) \; ,
\label{eq:2.1.1c}
\en
where the constant $\omega $ must be chosen to produce the
correct anomalous behavior under scale transformations.
The choice (\ref{eq:2.1.1c})  means that each instanton contributes
the same amount to the scale anomaly.
The averaging process over the instanton ensemble can now be performed
in a trivial manner. It is clear that the independent particle picture
is preserved during renormalization which is performed in the standard
way by rescaling $T^{a}_{\mu \nu } \rightarrow T^{a}_{\mu \nu } / Z_{T}$
corresponding to space-time independent counter terms.

Following the FSA,
we include  the determinant (\ref{eq:2.1.0}) in the action and
the generating functional for Green's function then becomes
\bea
Z &=& \int {\cal D}T \; \exp [ - S ]
\label{eq:2.1.1} \\
S[V] &=& \int d^{4}x \; \{ \frac{g^{2}}{4} T^{2} + \frac{i}{2}
T^{a}_{\mu \nu } F^{a}_{\mu \nu } [V]  \; + \;
\frac{ \mu_{0}^{4}(x) }{2} \tr \ln i\hat{T}/ \mu^{2} \} \; ,
\nonumber
\ena
where we have used the relation (\ref{eq:2.1.1a}).
$F^{a}_{\mu \nu }$ is the field strength functional (\ref{eq:2})
and $V^{a}_{\mu }[T]$ is the counterpart of the gauge potential in the FSA,
defined by equation (\ref{eq:10}).
We confine ourselves to the investigation of
instantons of the type (\ref{eq:a3}) whose field strength is given by
(\ref{eq:a2},\ref{eq:a17}).
As shown in section \ref{sec:2.1},
these instanton configurations minimize the first two terms of the
action in (\ref{eq:2.1.1}), giving rise to
the classical equation of motion (\ref{eq:11}).
For any solution to this equation,
in particular for the instantons of type (\ref{eq:a3}),
the effective action becomes
\be
S[V] \; = \; \int d^{4}x \;
\{ \frac{1}{4 g^{2} } F^{a}_{\mu \nu } [V] F^{a}_{\mu \nu } [V]
\; + \; \frac{ \mu_{0}^{4}(x) }{ 2 } \tr \ln \hat{F}[V]/ \mu^{2} \; \}
\label{eq:2.1.1d}
\en
and is a function of collective variables characterizing the instanton medium
(e.g., the instanton radius or the average distance of two instantons).
Substituting $\mu_{0}(x)$ from (\ref{eq:2.1.1c}),
the renormalized action becomes
\footnote{ Note that $F$ does not depend on the coupling strength. }
\be
S[V] \; = \; \int d^{4}x \;
\{ \frac{1}{4 g^{2} } F^{a}_{\mu \nu } [V] F^{a}_{\mu \nu } [V]
\; + \; \frac{ \omega }{ 2 } F^{2} \tr \ln \hat{F}[V]/ \mu^{2} \; \}
\label{eq:2.1.6}
\en
The dependence of this expression on the arbitrary scale $\mu $
can be removed by appealing to renormalization group arguments,
letting $g$ depend on $\mu $.
{}From $ dS / d \, \ln \mu =0 $ we obtain the $\beta $-function
\be
\beta (g) \; = \; - \, 8 (N^{2}-1) \omega \, g^{3} \; ,
\en
which, in fact, agrees with the asymptotic behavior of $\beta (g)$
in perturbative QCD.
So if the preexponential factor $\Det ^{-1/2} \hat{T}$ is the dominant
part of the determinant arising from fluctuations around an
instanton, we may set $\omega \approx \beta_{0}/8(N^{2}-1) $ with
$\beta _{0} = 11 N /48 \pi ^{2} $ for a pure SU(N) YM theory~\cite{yn83}.
This choice of $\omega $ also yields the correct scale anomaly. This
can easily be seen by examining the behavior of $S$
under a scale transformation, i.\ e.,
\be
\delta S \; = \; - 4(N^{2}-1) \omega \int d^{4}x \; F^{2}(x) \; .
\label{eq:2.1.6a}
\en
Here, the left hand side represents the trace of the energy-momentum tensor
$E^{\mu }_{\mu }$, which sets the fundamental energy scale in YM theories.
The choice $\omega = \beta_{0} / 8(N^{2}-1) $ then gives
\be
E^{\mu }_{\mu } = - \frac{ \beta _{0} }{2} F^{a}_{\mu \nu } F^{a}_{\mu \nu }
\; ,
\en
which is the correct result, first obtained by Collins et al.~\cite{co77}.

Inserting the one instanton solution (\ref{eq:a2}) into this
equation we obtain the action as a function of the instanton
radius $\rho $, i.\ e.,
\be
S_{\rho } \; = \; G^{2} \int d^{4}x \;
\{ (\frac{1}{4 g^{2} } + \frac{ \omega }{2} \ln \det \hat{G} ) \,
\psi ^{2} \; + \;
2 \omega (N^{2}-1)  \psi ^{2} \, \ln \psi /\mu^{2} \, \} \; ,
\label{eq:2.1.10}
\en
where $G^{2}= G^{a}_{\mu \nu } G^{a}_{\mu \nu } $ with the matrix
$G^{a}_{\mu \nu }$ defined in (\ref{eq:a2}).
A straightforward calculation yields the following
$\rho $-dependence
\be
S_{\rho } \; = \frac{ 8 \pi ^{2} G^{2} }{ 3 } [
\frac{1}{ 4 g^{2} } \; - \; \frac{ \beta _{0} }{2}  \, \ln (\rho \mu ) ] \; ,
\label{eq:2.1.6b}
\en
where we have inserted the explicit value for $\omega $ and
irrelevant numerical constants have been ignored.
\footnote{ Note that due to the presence of the explicit scale $\mu $,
the zero mode corresponding to scale transformations is absent,
implying that there are no further contributions from treating
this zero mode. }
Equation (\ref{eq:2.1.6b})  gives the same $\rho $-dependence
of the statistical weight  as obtained in the standard approach
if one includes the full contribution from the Gaussian fluctuations
around an instanton solution in the action~\cite{tho76,ber79}.
The difference, however, is that we have obtained this result
using physical arguments implying that $\beta _{0}$ is
undetermined at this level.
Note that large instanton radii are preferred here since
the coefficient of the logarithm in (\ref{eq:2.1.6b}) is negative.

Up to now, we have only considered  the weight
factor $\Det ^{-1/2} \hat{T}$. If we add the contribution from
the fluctuations of the field $T^{a}_{\mu \nu }$ to the effective
action, we expect that the only change will be in the value of $\omega $.
This is because we then recover the
standard instanton determinant calculated by
t'Hooft~\cite{tho76} which has precisely the $\rho $-dependence given
in (\ref{eq:2.1.6b}). This fact also implies
$\Det ^{-1/2} \hat{T}$ must already contain the dominant
quantum fluctuations relevant to instantons, and thus the tree level
field strength formulation gives the same $\rho $
dependence of the statistical weight as the full semiclassical
description in the standard YM approach.
In view of this, it is worthwhile to investigate the tree level
instanton interaction in field strength formulated QCD.
For this purpose we compare
the effective action (\ref{eq:2.1.6}) for two maximum spin
instantons localized at $x=0$ and $x=r$ and
with the same orientation, i.\ e.\ ,
\be
F^{a \, (2) }_{\mu \nu } \; = \;
G^{a}_{\mu \nu } \, [ \psi (x-r) + \psi (x) ] \; .
\label{eq:2.1.7}
\en
with the action of two independent instantons, giving
\be
S_{I} \; := \; S[ F^{(2)} ] \, - \, 2 S[F] \; .
\en
The first term
in (\ref{eq:2.1.6}) yields a repulsive interaction when the shape
of the instantons is kept fixed. Adding the second term, arising
from the integration over the gauge fields $A^{a}_{\mu}$, and thus
including the quantum effects of the standard approach, the potential
becomes short-range
repulsive and medium-range attractive
for $\rho < \rho _{c}$ but purely attractive for $\rho \ge \rho_{c}$.
The instanton medium is thus infra-red unstable.
The dependence of $S_{I}$
on the instanton distance in terms of the instanton radius
is shown in fig.\ (1) for different values $\rho \mu $
at fixed coupling strength $ \alpha = g^{2}(\mu ) / 4 \pi = 0.4 $.
Fig.\ (2) investigates the critical radius $\rho _{c}$
in terms of the coupling $g(\mu )$.
Assuming $\alpha = 0.4$ at a renormalization point $\mu \approx 1 \, $GeV
as suggested by perturbative QCD
the critical instanton radius is $\rho _{c} \approx 0.3 \, $fm.
It is seen that a medium of small
instantons ($\rho \mu < 1.48$) is stable for all values of $g(\mu )$.
On the other hand, there are always large instantons
(for fixed $g(\mu)$) which exceed the critical size $\rho _{c}$
and trigger a phase transition from an instanton gas to  an
homogeneous instanton medium.

{}From instanton liquid calculations~\cite{sh88}, it is  known
that the the vacuum distribution of the instanton radius is
sharply peaked. We have checked that for instantons with size larger
than some critical value there is no stabilization mechanism
preventing the classical instanton
vacuum from collapsing when $(\Det \hat{ T })^{-1/2}$ is
included in the classical approximation.
There are two major differences in our investigations as compared with
instanton liquid simulations; first,
in instanton liquid models quarks, play an important role
in the stabilization of the instanton medium, and we have neglected the
influence of quarks in our considerations so far. Second,
in instanton liquid models one usually includes only the
instanton-anti-instanton interaction, which is already present
at the tree level, while the instanton-instanton interaction,
which arises only from quantum fluctuations (and quarks),
is discarded. In our case,
order $\hbar$-effects yield the attractive instanton interaction
causing the infra-red instability.
Finally note that if one sticks to large but finite volume $L^{4}$,
implying that the instanton radius cannot exceed $L$, it is always
possible to choose a renormalization point at which
the instanton medium is stable.
\bigskip

\bigskip
\subsection{ Constant FSA solutions versus instantons }
\label{sec:3.2}
\medskip
In section \ref{sec:2.1} we saw that field strength
formulated YM theories of the form presented in \cite{ha77,re92},
with the weight factor $\Det ^{-1/2} \hat{T}$ excluded from
the effective action, have the same instanton solutions as exist in the
standard approach. We now investigate whether the field
strength approach (FSA) to YM-theories in the form proposed in \cite{sch90},
where the weight factor $\Det ^{-1/2} \hat{T}$ is included in the action
still allows for instanton-type solutions of the classical equation
of motion (\ref{eq:52}). Below we show that in the FSA, individual
localized instantons do not exist, as one might already expect
from the presence of an energy scale $\mu $ in the effective action.
For the SU(2) gauge group, the absence of the standard instantons
has already been proven in~\cite{sch90}.
For simplicity we restrict ourselves to instantons of the type
discussed in the previous section and use the ansatz (\ref{eq:a2}), i.\ e.\ ,
\be
g^{2} T^{a}_{\mu \nu }(x) \; = \; -i G^{a}_{\mu \nu } \, \psi (x^{2}) \; ,
\label{eq:53}
\en
where $G^{a}_{\mu \nu } $ is a constant
(anti-) self-dual matrix satisfying the
relation (\ref{eq:a6}). A surprising consequence of
the  property (\ref{eq:a6}) of the instanton
matrix $G^{a}_{\mu \nu }$ is that it implies (see appendix \ref{sec:b} )
\be
G^{a}_{\mu \nu } \; = \; \kappa f^{abc} ( \hat{G} ^{-1} )^{bc}_{\mu \nu }
\label{eq:55}
\en
where $\kappa $ is some constant depending on the underlying
gauge group and the dimension of space-time.
Setting
\be
T^{a}_{\mu \nu } \; = \; i \frac{ \mu ^{2} }{ g \sqrt{ \kappa } }
G^{a}_{\mu \nu }
\label{eq:58a}
\en
(\ref{eq:55}) becomes precisely the classical equation
of motion of the FSA
(\ref{eq:52}) for space-time independent field
configurations. Therefore, any instanton of the type considered in
section \ref{sec:3.1}, generates a homogeneous
solution (\ref{eq:58a}) of the FSA with the same color and
Lorentz structure. This is a major result of our paper.

Let us now go further and show that in the FSA these instantons
only generate (up to gauge rotations) space-time constant solutions.
For the configurations (\ref{eq:53}) the equation of motion (\ref{eq:52})
becomes
\be
G^{a}_{\mu \nu } \; [ \psi \, - \, \frac{ \mu^{4} g^{2} }{ \kappa }
\frac{1}{\psi }] \; - \;  F^{a}_{\mu \nu }[V] \; = \; 0 \; .
\label{eq:56}
\en
This equation only allows for constant $\psi (x^{2})$.
To see this explicitly we take the covariant derivative
$D^{ba}_{\mu } = \delta ^{ba} \partial _{\mu } + f^{bca} V^{c}_{\mu }$
of (\ref{eq:56}), giving
\be
G^{b}_{\mu \nu } \; \partial_{\mu } [ \psi \, - \,
\frac{ \mu^{4} g^{2} }{ \kappa } \frac{1}{\psi }] \; + \;
f^{bca} V^{c}_{\mu } G^{a}_{\mu \nu }
[ \psi \, - \, \frac{ \mu^{4} g^{2} }{ \kappa } \frac{1}{\psi }] \;
\; -  \;  D_{\mu }^{ba} F^{b}_{\mu \nu }[V] \;
= \; 0 \; .
\label{eq:57}
\en
By assumption $G^{a}_{\mu \nu }$ is (anti-) self-dual and by (\ref{eq:56})
the same must be true for $F^{a}_{\mu \nu }[V]$. The last term
then vanishes by the Bianchi identity
$D_{\mu } F^{\ast }_{\mu \nu }[V] =0 $.
Note, that any solution of (\ref{eq:56}) must also satisfy the
reduced equation (\ref{eq:57}), although the converse is not true.
Inserting the explicit form of $V^{a}_{\mu }$ (\ref{eq:10})
into (\ref{eq:57}) and using (\ref{eq:a6}), (\ref{eq:56})
may be simplified to
\be
G^{a}_{\nu \omega } x_{\omega } \, (\ln \psi)' \, \frac{1}{\psi }
\; = \; 0 \; .
\label{eq:58}
\en
This equation can only be satisfied by constant $\psi $.

We have thus shown that when $\Det ^{-1/2} \hat{T} $
is included in the effective tensor action,
the instantons disappear and instead constant, homogeneous
solutions with the same color and Lorentz tensor structure emerge.
In view of the results of the previous subsection
we may thus conclude that the homogeneous vacuum solutions of the
FSA represent coherent superpositions of uniformly oriented
instantons.

Note that the  condition (\ref{eq:a6})
on the matrix $G^{a}_{\mu \nu }$ of the instanton, is  more restrictive
than the equations of motion of the field strength approach (\ref{eq:52}).
Accordingly, the field strength approach allows further
solutions than those having the color and Lorentz structure
of the maximum spin instanton, discussed above.
For example, (\ref{eq:52}) allows solutions of mixed duality
(see section \ref{sec:3.3}). In the instanton approach such
configurations would correspond to a mixed
solid like medium of self-dual and antiself-dual instantons,
with different orientations and different localizations,
but which is found to be homogeneous after
averaging over all instanton positions and radii (but keeping color
and Lorentz structure fixed).

We conclude that the two instanton potential induced by
the additional weight factor $\Det ^{-1/2} \hat{T}$
in field strength formulated YM theory at tree level
is responsible
for the condensation of instantons, forming the of homogeneous
field strength vacuum seen in the FSA.

\bigskip
\section{ Instanton solids }
\label{sec:3.3}
\medskip
We have just seen that the $\tr \ln \hat{T}$ term
gives rise to a substantial instanton interaction and hence this term
deserves a non-perturbative treatment. One should therefore
include it in the action and thus in the definition of the
stationary point.
But this is  precisely the definition
the field strength approach (FSA) of ref.~\cite{sch90}
with the action (\ref{eq:51}).

We now go on to examine the spectrum of homogeneous solutions to the
classical equations of motion (\ref{eq:52}) of the FSA.
For constant $T^{a}_{\mu \nu }$ configurations these
reduce to
\be
g^{2} T^{a}_{\mu \nu } \; + \; \mu ^{4}
f^{abc} ( \hat{T}^{-1} )^{bc}_{\mu \nu } \; = \; 0 \; ,
\label{eq:62}
\en
which have the purely imaginary solutions~\cite{sch90}
\be
i g^{2} T^{a}_{\mu \nu } \; = \; G^{a}_{\mu \nu } \; .
\label{eq:60}
\en
These solutions describe a homogeneous vacuum with real
field strength $G^{a}_{\mu \nu }$
\footnote{ Also note that in view of (\ref{eq:11})
the instanton solutions carry real field strength
$
G^{2}_{Inst} (x) \; = \; (i g^{2} T_{Inst } )^{2} \; = \;
F^{a}_{\mu \nu } [V] \, F^{a}_{\mu \nu } [V] \; .
$ },
which minimizes the effective potential~\cite{sch90}
\be
\Gamma ( G ) \; = \; \frac{1}{4 g^{2} } G^{a}_{\mu \nu }
G^{a}_{\mu \nu }  \; - \; \frac{\mu^{4} }{2} \tr \ln (\hat{G}) \; .
\label{eq:61}
\en
{}From the equation of motion it follows that the constant solutions satisfy
the relation
\be
G^{a}_{\mu \nu } G^{a}_{\mu \nu } \; = \; 4 \, ( G^{a}_{i} G^{a}_{i}
+ \bar{G}^{a}_{i} \bar{G}^{a}_{i} ) \; = \; 4 (N^{2}-1)
\mu ^{4} \, g^{2} \; .
\en
If we consider the expansion coefficients $G^{a}_{i}$ and
$\bar{G}^{a}_{i}$ as cartesian coordinates of $R^{6(N^{2}-1)}$,
then the classical solutions are given by points on the
unit sphere $S^{6 (N^{2}-1)-1}$.

In ref.~\cite{sch90} the various classical solutions have been
classified according to the topological quantum number
\be
m \; = \; \frac{1}{2} ( N_{+} - N_{-}) \; ,
\label{eq:3.4.1}
\en
where $N_{\pm }$ are the number of positive and negative eigenvalues
of the matrix $\hat{G}$. Solutions with different $m$ are
separated by infinite energy barriers, since $\tr \ln \hat{G}$
diverges when one of the eigenvalues crosses zero.

\bigskip
\subsection{ Lowest action solution of the FSA }
\label{sec:3.4.1}
\medskip
It was shown in section \ref{sec:3.2} that any instanton
solution of the type discussed in section \ref{sec:3.1}
provides a homogeneous (anti-) self-dual solution to the
equation of motion (\ref{eq:52}) of the FSA.
Surprisingly, it is just the maximal spin $(s=(N-1)/2)$ instanton of the
SU(N) YM-theory that provides the lowest action solution of the FSA.
More precisely, in SU(2) (in four space-time dimensions)
it has been shown there exist only (up to gauge and Lorentz transformations)
six independent constant solutions, which are all degenerate.
Four of them are (anti-) self-dual solutions arising from the maximal spin
$s=1/2$ SU(2) instantons. These solutions carry the topological
quantum number (\ref{eq:3.4.1}) $m=\pm 2$. There are two further
solutions which are not (anti-) self-dual, carry $m=0$,
and are only accidentally degenerate with the self-dual solutions.
The lowest action solution for SU(3), the maximal spin instanton $(s=1)$,
which has $m=4$, is non-degenerate (up to symmetry transformations).

In order to get a better understanding  of the homogeneous solutions,
generated by instantons, of the FSA equation of motion we
investigate the structure of the matrix $\hat{G}$.
For self-dual field configurations we have
\be
\hat{G}^{ab}_{\mu \nu } \; = \; \hat{G}^{ab}_{i} \eta ^{i}_{\mu \nu }
\; \hbo \hat{G}^{ab}_{i} \; = \; f^{abc} G^{c}_{i} \; .
\label{eq:3.4.2}
\en
For the classical FSA solutions which are generated by instantons,
the $G_{k}$ (\ref{eq:18i}) defined in the fundamental representation
of the SU(N) gauge group, form an SU(2) algebra (see eq.(\ref{eq:34}))
in some spin $s$ representation. The same must be true for
the matrices $-i \hat{G}_{k}$ which are members of the adjoint
representation of SU(N). These form an (in general reducible)
SU(2) representation, i.\ e.\ ,
\be
L_{k} := -i \hat{G}_{k} \; \hbo [ L_{k}, L_{i} ] \; = \;
i \epsilon _{kim} L_{m} \; ,
\label{eq:3.4.3}
\en
with different spin $L \not= s$.
Furthermore, the $S^{k} := \frac{i}{2} \eta ^{k}$
form the four dimensional $s=1/2$ representation of SU(2).
Hence for these solutions the matrix $\hat{G}$
can be represented entirely in terms of SU(2) generators, i.\ e.\ ,
\be
\hat{G}^{ab}_{\mu \nu } \; = \; 2 \, \vec{L} _{ab} \vec{S}_{\mu \nu } \; .
\label{eq:3.4.5}
\en
Due to the absence of anti self-dual components, i.e.\
of the spin $\bar{S}^{k}= \frac{i}{2} \bar{ \eta }^{k}$ all
eigenvalues of $\hat{G}$ are (at least) two-fold degenerate.
Introducing  the grand spin $\vec{K} = \vec{L} + \vec{S} $, the matrix
$\hat{G}$ becomes
\be
\hat{G} \; = \; \vec{K} ^{2} \, - \, \vec{L}^{2} \, - \, \vec{S}^{2}
\label{eq:3.4.6}
\en
and its eigenvalues are
\be
\lambda \; = \; k(k+1) \, - \, l(l+1) \, - \, s(s+1) \; ,
\label{eq:3.4.7}
\en
where $k=l \pm \frac{1}{2}$. If $\vec{L}$ is an irreducible
representation of SU(2) the
eigenvalues $\lambda $ are $(2k+1)(2 \bar{s}+1)$ degenerate
$(\bar{s}=1/2)$.
If $\vec{L}$ decays in $\nu $ irreducible SU(2) representations,
the degree of degeneracy is
$(2 \bar{s}+1) \sum_{\rho =1 }^{\nu }  (2 k_{(\rho)}+1)$.
Since the adjoint representation of SU(N) is $(N^{2}-1)$-dimensional,
we have the constraint
\be
2 \, \sum_{\rho =1}^{\nu }
\sum_{k_{(\rho )} =l_{(\rho )} \pm 1/2 } (2 k_{(\nu )} +1) \; = \;
4 \, \sum_{\rho=1}^{\nu }  (2 L_{(\rho )}+1) \; = \; 4 \, (N^{2}-1) \; .
\label{eq:3.4.8}
\en
Furthermore there are
\be
N_{+} \; = \; 4 \, \sum_{\rho=1 }^{\nu } (L_{(\rho )} +1)
\label{eq:3.4.9}
\en
positive eigenvalues $(k=l+1/2)$ and
\be
N_{-} \; = \; 4 \, \sum_{\rho=1}^{\nu } L_{(\rho )}
\label{eq:3.4.10}
\en
negative eigenvalues $(k=l-1/2)$ of $\hat{G}$, so that the
topological quantum number (\ref{eq:3.4.1}) becomes
\be
m \; = \; \frac{1}{2} (N_{+}-N_{-}) \; = \; 2 \, \nu \; .
\label{eq:3.4.11}
\en
Since the number of irreducible representations is restricted
by (\ref{eq:3.4.8}) their is a maximum topological quantum number for
each SU(N) group. In the case of SU(2) the adjoint representation
is 3-dimensional and can accommodate only a single spin
representation ($L=1/2$ or $L=1$, the latter is realized for the
maximum spin solution).
Therefore $\nu =1$
and the maximal quantum number is $m=2$, in agreement with the
explicit solutions~\cite{sch90}. In the case of SU(3) the adjoint
representation is eight dimensional and,
for the maximum spin solution, may be decomposed into
$L_{(1)}=1$ and $L_{(2)}=2$ irreducible SU(2) representations
implying $m \le 4$, in agreement with numerical simulations
(see section \ref{sec:3.4.2}). It seems that this property
is not restricted to the instanton like solutions, since up to
now no configuration $\hat{G}$ with $m > 4$ has been found numerically by
choosing $G^{a}_{\mu \nu }$ randomly.

\bigskip
\subsection{ Statistical weight of the constant vacuum solutions }
\label{sec:3.4.2}
\medskip
In this section we investigate the weight with which
each classical solution contributes to the path integral.
We argue that the completely
self-dual FSA solutions discussed in the previous subsection,
should be statistically suppressed compared to the solutions
with mixed duality. The reason is that the self-dual solutions
can be interpreted as a coherent superposition of maximum
spin instantons with unique orientation, whilst the mixed
duality solutions represent superpositions of instantons
and anti-instantons.

In the interesting case of an SU(3) gauge group,
many classical solutions are known to exist. It is therefore of great
practical interest to investigate which are
the most important ones dominating the quark interaction
(see (\ref{eq:13})). For this purpose we assume
that the inclusion of the quarks modifies the gluonic sector
only slightly, a fact which seems to be supported by the QCD sum
rules~\cite{no84}.

Each classical solution contributes a weight factor
$\exp [- \Gamma (G_{c}) / \mu ^{4} ]$ with $\Gamma (G)$
defined by (\ref{eq:61}). Upon rescaling
$T^{a}_{\mu \nu } \rightarrow T^{a}_{\mu \nu } / g$ one can explicitly
check, using (\ref{eq:60}) and (\ref{eq:61}), that this weight
factor is independent of $g$ and $\mu $ and therefore renormalization
group invariant. A second weight factor arises from the integration
over the Gaussian fluctuations in the tensor field
around a classical solution. These weight factors have been evaluated
for SU(2) in three dimensions~\cite{re91}. For SU(3) in four dimensions
however, the evaluation of these weight factors for all
classical vacuum solutions is not feasible.
We therefore take an alternative approach, motivated by the
stochastical quantization method.
This approach is not only feasible for SU(3) but also
allows a deeper insight into the phase space structure of the FSA.

We generate a series of field strengths by the iteration
\be
G^{a \, (n+1) }_{\mu \nu } \; = \; G^{a \, (n) }_{\mu \nu }
\; - \; \tau \frac{ \partial \Gamma }{ \partial G^{a}_{\mu \nu } }
[ G^{(n)} ] \; + \; \chi ^{a \, (n) }_{\mu \nu } \; ,
\label{eq:64}
\en
where $\chi ^{a \, (n)}_{\mu \nu }$ are random numbers with a Gaussian
distribution. From Parisi's work~\cite{pa81}
it is well known that the
$G^{(n)}$ are distributed according $\exp [- \Gamma (G) ]$ if
the width of the $\chi $-distribution, $w$ and $\tau $
are related by $w=\sqrt{4 \tau }$.
Putting $\chi ^{a \, (n)}_{\mu \nu } =0$, the iteration
(\ref{eq:64}) reduces  to a fixed point
equation and the $G^{a \, (n)}_{\mu \nu }$ converge to the classical value
$G^{a}_{\mu \nu \, c}$ in the course of the iteration:
\be
G^{a \, (n+1) }_{\mu \nu } \; = \; G^{a \, (n) }_{\mu \nu }
\; - \; \tau \frac{ \partial \Gamma }{ \partial G^{a}_{\mu \nu } }
[ G^{(n)} ] \; , \hbo ( \tau \leq 1 ) \; .
\label{eq:65}
\en
In order to go beyond the classical approximation, we decompose
the $G$-space that is integrated over in the generating functional
for gluonic Green's functions
\be
Z_{G} \; = \; \int {\cal D} G \; \exp [ - \Gamma (G) ]
\label{eq:66}
\en
into domains $D_{i}$.
Each domain is characterized by a single classical solution
and all the iterations (\ref{eq:65}) with a starting point inside
such a domain converge to this configuration. We therefore write
\be
Z_{G} \; = \; \sum_{i} \int _{ \{ D_{i} \} } {\cal D} G \;
\exp [ - \Gamma (G) ] \; ,
\label{eq:67}
\en
where $\{D_{i}\}$ indicates that the integration extends only over
the domain $D_{i}$.
Performing a classical approximation in each domain we obtain
\be
Z_{G} \; \simeq \; \sum_{i} \exp [ - \Gamma ( G_{c}^{(i)} ) ] \;
\int _{ \{ D_{i} \} } {\cal D} G \; .
\label{eq:68}
\en
This implies that the additional weight factor is given by the
volume of a domain corresponding to an attractive fixed point
(classical solution).
These domains posses a highly non-trivial,
non-continuous structure. However, their volumes can be
estimated by randomly choosing a starting configuration
$G^{a}_{\mu \nu }$ and observing which solution the
iteration (\ref{eq:65}) converges to. The result of 100000 sweeps is
presented in table \ref{tab:1}. The first column gives the classical
action of a particular solution and the second column
gives the quantum number $m$ defined by (\ref{eq:3.4.1}).
The third column lists the number of events a
particular solution attained within 100000 sweeps. Also given
is the effective action including the statistical weight factor:
\be
S_{eff} \; = \; \sum_{i} \bigl[ \Gamma ( G_{c}^{(i)} ) \; - \;
\ln \int _{ \{ D_{i} \} } {\cal D} G \, \bigr] \; .
\label{eq:69}
\en
The spectrum of solutions classified by the effective action and
m-sector quantum number is shown in figure (3).
\begin{table}[t]
\begin{tabular}{|cclc|} \hline
$S_{c}$  &  m-sector  & events & \hspace{5mm} $S_{eff}$ \hspace{5mm}
\\ \cline{1-4}
1.3041   &  4         & 15     & 8.3943  \\
2.1836   &  0         & 18003  & 2.1836  \\
2.2714   &  2         & 6911   & 3.2288  \\
2.4751   &  1         & 7463   & 3.3556  \\
2.5421   &  2         & 433    & 6.2697  \\
2.6382   &  0         & 16377  & 2.7779  \\
2.6736   &  0         & 9290   & 7.1434  \\
2.7402   &  0         & 9090   & 3.4236  \\
2.7422   &  1         & 4414   & 4.1480  \\
2.7662   &  0         & 6116   & 3.8458  \\
2.7946   &  0         & 3435   & 4.4511  \\
2.8039   &  1         & 10927  & 3.3032  \\
2.8076   &  1         & 2736   & 4.6916  \\
2.9978   &  0         & 1606   & 5.4146  \\
3.0118   &  2         & 66     & 8.6204  \\
3.0285   &  0         & 122    & 8.0228  \\
3.0302   &  0         & 61     & 8.7176  \\
3.1539   &  1         & 1545   & 5.6094  \\
3.1834   &  0         & 1347   & 5.7760  \\
3.2335   &  2         & 43     & 9.2706  \\ \hline
\end{tabular}
\label{tab:1}
\end{table}
The classical homogeneous field strength solution corresponding
to the maximal spin $s=1$ self-dual SU(3) instanton (\ref{eq:27i})
carries $m=4$.
This solution is the only self-dual SU(3) field strength solution
found so far and has the lowest classical action
$\Gamma (G_{c})=1.3041$~\cite{al91}.
However, it is
strongly statistically suppressed as expected since the configurations
of mixed duality can be formed in many more ways
by superposing instantons and anti-instantons.
The solution with the lowest effective action is an $m=0$ configuration
with $S_{eff}=2.1836$ found in~\cite{sch90}.
The solution is eightfold degenerated and four of the orientations
of the chromo-electric and magnetic fields are shown in figure (4).
The remaining four configurations are obtained from figure (4) by
reflecting the magnetic field $B^{a} \rightarrow - B^{a}$.

\bigskip
\section{ Quark condensation in the gluonic background }
\label{sec:4}
\medskip
So far, we have mainly confined ourselves to the description of
the ground state of pure Yang-Mills theory. We now turn to the
investigation of the quarks in this gluonic vacuum. The quark ground
state directly influences the low energy hadron physics. It is
generally believed that the $SU(n)_{L} \times SU(n)_{R}$ chiral (flavour)
symmetry is spontaneously broken to the diagonal $SU(n)_{V}$
flavour group by quark-condensation.
A similar mechanism of spontaneous symmetry breaking was first
proposed by Nambu and Jona-Lasinio in a model
with a chiral invariant four fermion contact interaction~\cite{na61}.
A more realistic model, based on QCD, was proposed by Callan et
al.\ \cite{ca78}. In this model a four-fermion interaction is
induced by instanton anti-instanton interactions
which leads to quark condensates of the right order
of magnitude.

In this section we first
investigate the quarks in an isolated instanton background and
take the average over instanton positions and sizes.
The results obtained are then compared
with those of the field strength approach, where the
quarks move in a vacuum of constant field strength representing a
condensate of instantons.
We argue that the averaging process over the instanton ensemble
yields the same gap equation for the quark condensates as in the
field strength approach where the averaging is automatically
performed by including the weight factor in the action
(see section \ref{sec:2.2}).

\bigskip
\subsection{ Quarks in an  isolated instanton field }
\label{eq:4.1}
\medskip
We wish to investigate the quark theory obtained in (\ref{eq:12}).
To do so we assume that instantons of the pure Yang-Mills theory
dominate the $T$-functional integral and treat the additional
pre-exponential factor
$\Det ^{-1/2} \hat{T} $ as weight factor for each classical
configuration. Identifying the external source with the quark
color octet current $j^{a}_{\mu } = \bar{q} \gamma _{\mu } t^{a} q$,
the generating functional of the quark theory (\ref{eq:12},\ref{eq:13})
in the lowest order stationary phase approximation
$T^{a}_{\mu \nu } = - \frac{i}{g^{2}} F^{a}_{\mu \nu }[V] $
(\ref{eq:11x}) becomes
\bea
W &=& \int {\cal D} q \; {\cal D} \bar{q} \; \Det ^{-1/2}
\hat{F}[V] \; {\cal G} [j] \; \exp [ - \int d^{4}x \;
\bar{q}( i \dslash + im) q \, ]
\label{eq:14} \\
{\cal G} [j] &=& \exp [ \int d^{4}x \; \{ j^{a}_{\mu } V^{a}_{\mu }
+ \frac{ g^{2} }{2} j^{a}_{\mu }(x) ( \hat{F} ^{-1}[V](x) ) ^{ab}_{\mu \nu }
j^{b}_{\nu }(x) \} \, ] \; .
\label{eq:15}
\ena
Since the field strength of an instanton vanishes asymptotically,
the matrix $\hat{F}^{-1}[V](x)$ diverges for large $x$, implying
that non-zero color currents $j^{a}_{\mu } \not= 0$
are localized near the instanton position.
To get a feel for the quark ground state we investigate the
quarks in a one instanton background field, i.\ e.\ ,
\be
F^{a}_{\mu \nu } [V] \; = \; G^{a}_{\mu \nu } \frac{ 4 \rho ^{2} }{
( x^{2} + \rho ^{2} )^{2} } \; = \; G^{a}_{i} \eta ^{i}_{\mu \nu }
\, \psi(x^{2}) \; .
\label{eq:16}
\en
The use of Feynman's variational principle $ \langle e^{F}
\rangle \ge e^{ \langle F \rangle } $
where the angle brackets denote averaging over the collective
coordinates of the instanton,
implies that the bilinear term in
the quark current is more important than the linear term
since $ \langle j^{a}_{\mu } \rangle =0$. As an initial estimate
we skip the linear term in $j^{a}_{\mu }$. This
should in fact be a good approximation if one calculates static ground state
properties like condensates, but will probably fail if dynamical properties
like meson masses are required. Adopting the usual bosonization
procedure, introducing mesonic fields, we insert the identity
\bea
1 &=& \int {\cal D} \Omega \; \delta [ \Omega ^{ab}_{ik} -
i q^{a}_{i}(x) \bar{q}^{b}_{k}(x) ]
\label{eq:18} \\
&\sim &
\int {\cal D} \Omega \; {\cal D} \Sigma \; \exp [- i \int d^{4}x \;
\tr \, \Sigma ( \Omega -i q \bar{q} )]
\nonumber
\ena
into (\ref{eq:14}) and integrate over the quark fields. The classical
equations of motion for the $\Sigma $ and $\Omega $ fields
are the Dyson-Schwinger equations. In the strong coupling
approximation to the quark propagator~\cite{la91a,la91b} they are
\bea
i \Sigma (x) &=&  \mu _{0}(x) ^{4} \, \Omega ^{-1}(x) \; ,
\label{eq:19} \\
i \Sigma (x) - m &=&  g^{2} \gamma _{\mu } t^{a} \, \Omega (x) \,
\gamma _{\nu } t^{b} \, (\hat{F}^{-1}_{B}) ^{ab}_{\mu \nu }(x)
\label{eq:20}
\ena
where $\mu _{0}(x)$ is the space-time dependent regulator (
introduced in section \ref{sec:3.2a}).
For large current masses $m$ the solution of this system
of equations is
\be
\Omega (x) \; = \; \frac{ \mu_{0} ^{4}(x) }{ m } \; .
\label{eq:20a}
\en
In order not to spoil the independent pseudo-particle picture
(compare section \ref{sec:3.2a}) we
impose the asymptotic relation between the quark- and
gluon-condensate~\cite{sh79},
$ -i m \langle :\bar{q}q: \rangle \; = \;
\frac{1}{12} \langle : \frac{ \alpha_{s} }{ \pi } GG : \rangle $,
even with one instanton. This yields with
$-i \langle :\bar{q}q: \rangle  = \tr \Omega $ and, using (\ref{eq:20a}),
\be
\mu_{0}^{4} (x) \; = \; \frac{1}{144} \frac{1}{4 \pi^{2} } F^{2} \; .
\label{eq:20b}
\en
Up to an irrelevant constant, this is the same relation between
the local regulator $\mu_{0}(x)$ and the field strength
of the instanton $F^{a}_{\mu \nu }(x)$ found in section
\ref{sec:3.2a} (see eq.(\ref{eq:2.1.1c})).
Again the divergences of $\mu_{0}(x)$
can then be absorbed by a field renormalization.

In the chiral limit $(m=0)$
an explicit solution of (\ref{eq:20a}) and
(\ref{eq:20b}) in the one instanton background,  with $F^{a}_{\mu \nu }$
given by (\ref{eq:16}), is
\be
\Omega (x) \; = \; \Omega _{0} \, \frac{ \mu _{0}(x) ^{2} }{g}
\sqrt{ \psi (x^{2}) }
\label{eq:21}
\en
where the constant matrix $\Omega _{0}$ satisfies the algebraic equation:
\be
\Omega ^{-1}_{0} \; - \; \gamma _{\mu } t^{a} \, \Omega _{0} \,
\gamma _{\nu } t^{b} \, (\hat{G}^{-1}) ^{ab}_{\mu \nu }(x) \; = \; 0 \; ,
\label{eq:22}
\en
with $G^{a}_{\mu \nu }$ being the instanton matrix in (\ref{eq:a2}).
The contribution of the quarks localized in
a single instanton (\ref{eq:16}) to the total condensate is
\be
\int d^{4}x \; \langle i \bar{q} q \rangle _
{I} \; = \; - \frac{ \sqrt{ G^{2} } }{ 24 \pi g }
 \, \tr \Omega _{0} \, \int d^{4}x \; \psi ^{3/2} (x^{2})  \; .
\label{eq:23}
\en
The subscript $I$ indicates that it is only the contribution
of one instanton that is considered.
A straightforward evaluation of the integral for the
profile function (\ref{eq:16}) yields
\be
- \, \langle im \bar{q}q \rangle _{I} \; = \;
\frac{ \sqrt{ G^{2} } \pi }{ 6 } \, \frac{m}{g} \,
\rho \, \tr \Omega _{0}
\label{eq:24}
\en
The ratio of the quark- and gluon-condensate for the one instanton
configuration is thus
\be
- \frac{ \langle im \bar{q}q \rangle }
{ \langle \frac{ \alpha_{s} }{ \pi } GG \rangle } \; = \;
\frac{ \pi }{ 4 \sqrt{ G^{2} } } \, \frac{m}{g} \rho \,
\tr \Omega _{0} \; .
\label{eq:24a}
\en
Since we have preserved the independent instanton picture
during renormalization the total condensates are obtained by multiplying by
the instanton density. This implies that the ratio (\ref{eq:24a}) already
displays the total value. However, the ratio is divergent
if there is no stabilization of the instanton at a certain radius.
This indicates that the quarks might play an important role
in stabilizing the instanton medium.
Since the quark condensate is a renormalization
group invariant we observe that the renormalized current mass
scales with $g$, which is the approximate behavior
known from perturbative QCD.

\bigskip
\subsection{ Quark condensation in the field strength approach }
\label{sec:4.2}
\medskip
As discussed in section \ref{sec:3.3}, the field strength approach
provides a non-trivial gluonic vacuum with a non-vanishing
gluon condensate. Furthermore, the non-vanishing classical field strength
mediates a Nambu-Jona-Lasinio type of quark interaction. This
quark interaction has been investigated in the
literature for the gauge group SU(2) analytically in ref.~\cite{re91}
and for SU(3) numerically in refs.~\cite{la91a,al92}.
Using the gluonic configuration with
the lowest effective action from section \ref{sec:3.3} (shown in
figure (4), the quantitative results for the condensates agree
with the experimental values~\cite{la91a}. Here we briefly review the
mechanism leading to quark condensation in the field strength approach
in order to compare it with the quark condensation in the
instanton approach discussed in the previous section.

The generating functional for quark Green's functions in the FSA is obtained
from (\ref{eq:50}, \ref{eq:51}) and (\ref{eq:13}). Assuming that
the gluonic integral is dominated by a classical configuration
$g^{2} T^{a}_{\mu \nu }= -i G^{a}_{\mu \nu } $ which are constant in
a certain color and Lorentz frame,
the Euclidean generating functional becomes
\be
W = \int {\cal D} q \; {\cal D} \bar{q} \; \exp [
- \int d^{4}x \; \{ \bar{q} (i \dslash + im) q
\, - \, \frac{ g^{2} }{2} j^{a}_{\mu }
(\hat{G}^{-1})^{ab}_{\mu \nu } j^{b}_{\nu } \; ,
\label{eq:80}
\en
where we have used the fact that $V^{a}_{\mu }=0$ for (in this frame)
homogeneous field strength configurations. Note that an average over
all gauge and Lorentz equivalent classical configurations
is understood in (\ref{eq:80}).
We apply the
same bosonization procedure as in the previous section,
introducing the condensate variable $\Omega $, and meson fields $\Sigma $.
The classical condensate $\langle i:\bar{q}q: \rangle = \tr \Omega _{c}$ in
the zero momentum approximation (for details see \cite{la91a})
can be calculated from the gap-equation, i.\ e.\ ,
\be
\mu ^{4} \Omega _{c}^{-1} \; - \;
t^{a} \gamma _{\mu } \, \Omega _{c} \, t^{b} \gamma _{\nu }  \;
g^{2} (\hat{G}_{c}^{-1})^{ab}_{\mu \nu } \; = \; 0
\label{eq:81}
\en
where $\mu $ is a momentum cutoff, rendered finite by a renormalization
of the quark fields. This equation is the counterpart of (\ref{eq:22})
of the instanton picture in the field strength formulation
and obviously exhibits the same
algebraic structure. However, the renormalization procedure in the
field-strength approach is greatly simplified, because we only have
constant field strength configurations and so no averaging process
is required. We only need to absorb the divergence
of the momentum cutoff by redefining the quark fields.
For numerical details we refer the reader to \cite{la91a,la91b}.

\bigskip
\section{ Concluding remarks }
\label{sec:7}
\medskip
An alternative formulation of YM theories can be made in terms
of field strengths.
Recently it was shown~\cite{re92} that this field strength formulation yields
the same semiclassical description as the conventional formulation in terms
of gauge fields. Not only the same classical
solutions (instantons) are obtained, but also the quantum
fluctuations around the instantons give the same result in both
approaches. The field strength formulation is, however,
superior to the standard approach at the tree level
since the corresponding quantum transition amplitude
already includes quantum effects.
In the standard approach these will only show up beyond the tree level.
These quantum corrections give rise to an instanton-instanton
interaction and are contained in a functional determinant
arising from integrating out the gauge potential in the field
strength formulation. In the FSA this functional determinant
is included into an effective action of the field strengths.
This effective action does not admit explicit instantons as stationary
points but rather homogeneous field strength configurations.
We have presented a class of SU(N)
instantons, the maximum spin instantons, which
give rise to the vacuum solution of lowest classical action in the FSA.
These constant field strength solutions mediate a quark interaction which
leads to quark condensation very similarly to the instanton induced
quark interaction proposed by Callan et al.

We have further shown that there is an intrigueing connection
between the FSA and the more traditional instanton physics.
In particular our investigations have revealed that the homogeneous
vacuum of the FSA can be interpreted as a solid of condensed
instantons aligned in color and Lorentz space.
In the instanton liquid approach, fermions (and in principle,
quantum fluctuations of the gauge field too) give rise to strong instanton
correlations leading to a liquid like structure of the
instanton vacuum. Recent investigations~\cite{ver89}, where
shape variations of instantons are included, indicate that the
strongly correlated instantons might loose their identity tending
to form a homogeneous ground state.
This is precisely what one finds in the FSA, where one obtains
homogeneous field strength configurations as a stationary solutions.
Thus if sufficient correlations are included, the instanton picture
might lead to a similar vacuum as the FSA. However, the FSA
provides a more efficient description of the non-perturbative vacuum
than the instanton approach. While in the latter case higher order
correlations have to be included, in the FSA a strongly correlated vacuum
is already obtained at tree level, which already includes a good
deal of of the higher order
quantum fluctuations of the conventional formulation.

\bigskip
\leftline{ \bf Acknowledgements: }
\medskip
The authors are grateful to F.\ Langbein for a careful reading
of the manuscript and critical remarks.

\appendix
\bigskip
\section{Instanton induced classical field strength solutions}
\label{sec:b}
\medskip
We have to show that the instanton matrix $G^{a}_{\mu \nu }$
defined by either
(\ref{eq:a6}), (\ref{eq:32}) or (\ref{eq:34}) also satisfies
\be
G^{a}_{\mu \nu } \; = \; \kappa \, f^{abc}
(\hat{G}^{-1}) ^{bc}_{\mu \nu }
\label{eq:b1}
\en
where $\kappa $ is some constant. The crucial step is a  calculation
of  the inverse of $\hat{G}^{ab}_{\mu \nu }= f^{abc} G^{c}_{\mu \nu }$.
For this end we note that for instanton solutions the matrices
\be
L_{k} = -i \hat{G}_{k} = G^{a}_{k} T^{a} \; , \hbox to 2cm {\hfil and \hfil }
(T^{a})_{bc}= -i f^{abc}
\en
define an SU(2) algebra as we have already seen in section
\ref{sec:3.4.1}, i.\ e.\ ,
\be
[ L_{k} , L_{l} ] \, = \, i \epsilon _{klm }  L_{m} \; .
\label{eq:b1a}
\en
For SU($N>2$) gauge groups, this SU(2) representation is, however,
reducible. But since the matrix $\hat{G}$ is assumed to be regular,
the $\hat{G}_{i}$ must be built up entirely from spin
$L^{(k)} \not= 0, \; (k=1,2,\ldots )$ irreducible representations.
In particular, the $-i \hat{G}_{k}$ can be expressed in the form
\be
-i \hat{G}_{k} \; = \; \left(
   \begin{array}{cccc}
      L^{(1)}_{k}  &             &           &             \\
                   & L^{(2)}_{k} &           &             \\
                   &             & \ddots    &             \\
                   &             &           & L^{(r)}_{k} \\
   \end{array} \right)
\label{eq:**}
\en
where the $L^{(i)}_{k}$ denote the $(2 L^{(i)}_{k} +1)$-dimensional
irreducible representations. Since the $\hat{G}_{k}$
have dimension $(N^{2}-1)$ we must have
\be
\sum_{p=1}^{r} (2 L^{(p)}_{k}+1) \; = \; N^{2}-1 \; .
\en
Furthermore, since the $\hat{G}_{k}$ are antisymmetric matrices,
only antisymmetric spin representations can enter (\ref{eq:**}).
For the gauge group SU(2) (where only the maximum spin
instanton exist) the $\hat{G}_{k}$ are three dimensional and must thus be
given by the spin $L=1$ irreducible representation, i.\ e.\ ,
\be
(\hat{G}_{k})^{ab} \; = \; \epsilon ^{kab} \; .
\en
For SU(3) the matrices $\hat{G}_{k}$ are eight dimensional
and can be built up either from the two four dimensional
antisymmetric spin $s=1/2$ representations
($\eta ^{k} $ or, $\bar{\eta }^{k}$) or from
an $L=1$ and an $L=2$ representation. For the maximum spin
$s=(N-1)/2=1$ instanton, it turns out that the latter case is realized.
In order to find the inverse of $\hat{G}$ we calculate
\be
\hat{G}^{ac}_{\mu \alpha } \hat{G}^{cb}_{\alpha \nu } =
\hat{G}^{ac}_{i} \hat{G}^{cb}_{l} \, \bigl( \frac{1}{2}
\{ \eta^{i}, \eta ^{l} \} + \frac{1}{2} [ \eta^{i}, \eta^{l} ]
\bigr) = M^{ab} \delta _{\mu \nu } + \hat{G}^{ab}_{\mu \nu } \; ,
\label{eq:b2}
\en
where we have used the algebra of the t'Hooft symbols (\ref{eq:15i})
and the instanton properties (\ref{eq:32}, \ref{eq:b1a}).
The matrix $M$ is the SU(2) Casimir operator in the
adjoint representation of SU(N), i.\ e.\ ,
\be
M := L_{k} L_{k} \; ,
\en
which is symmetric by construction.
{}From the representation (\ref{eq:**}) it is clear that this matrix
is given by the direct sum of the quadratic Casimir operators
of the irreducible SU(2) representations $L^{(i)}_{k}$
\be
M \; = \; \left(
   \begin{array}{ccc}
      L^{(1)} ( L^{(1)}+1 ) \; 1_{2 L^{(1)}+1 } &   &      \\
         &   L^{(2)} ( L^{(2)}+1 ) \; 1_{2 L^{(2)}+1 }  &   \\
         &   &  \ddots   \\
   \end{array} \right) \; .
\label{eq:***}
\en
The matrix $M$ is hence regular and commutes with $\hat{G}$
from (\ref{eq:**}). We thus find that the inverse of $\hat{G}$ is given by
\be
(\hat{G}^{-1})^{ab}_{\mu \nu } \; = \; (M^{-1})^{ac} \,
[ \hat{G} ^{cb}_{\mu \nu } - \delta ^{cb} \delta _{\mu \nu } ] \; .
\label{eq:b6}
\en
For the FSA equation of motion, we only require the projection of
$\hat{G}^{-1}$ onto the SU(N) adjoint representation:
\be
f^{cab} (\hat{G}^{-1})^{ab}_{\mu \nu } \; = \; - \,
\tr \{ T^{c} \, M^{-1} \, i \hat{G}_{l} \} \, \eta^{l} _{\mu \nu } \; .
\label{eq:b7}
\en
Note that there is no contribution from the unit matrix on the right hand side
of (\ref{eq:b6}) since the matrix $M$ is symmetric.
If the SU(2) representation $\hat{G}_{k}$ is irreducible the Casimir
operator $M$ (as well as its inverse) is proportional to the unit matrix.
In this case, (\ref{eq:b7}) directly implies that the equations
of motion of the FSA (\ref{eq:b1}) are satisfied.
Straightforward algebraic manipulation shows that
\be
M^{ab} G^{b}_{l} \; = \; - f^{ame} G^{e}_{k} f^{mbd} G^{d}_{k}
G^{b}_{l} \; = \; 2 \, G^{a}_{l} = L(L+1) G^{a}_{l}
\; \hbox to 1.5 true cm{ \hfill for }
\; \; \; \; l=1 \ldots 3 \; ,
\label{eq:116}
\en
which implies that the $L_{k}$ always
contain a spin $L=1$ representation.
The adjoint representation of SU(2) is 3-dimensional
and hence irreducible. However, (\ref{eq:116})  also implies that the
representation $L_{k}$  is reducible for SU($N>2$) if
we insist that the inverse matrix $\hat{G}^{-1}$ exists $(\det M \not= 0)$.
In fact as we have already seen for SU(3), $L_{k}$ must be given by the
direct product of spin $L=1$ and $L=2$ representations.
{}From (\ref{eq:***}) we find for the inverse
\be
M^{-1} \; = \; \left(
   \begin{array}{ccc}
      \frac{1}{ L^{(1)} ( L^{(1)}+1 )} \, 1_{2 L^{(1)}+1 } &   &      \\
         &   \frac{1}{ L^{(2)} ( L^{(2)}+1 )} \, 1_{2 L^{(2)}+1 }  &   \\
         &   &  \ddots   \\
   \end{array} \right) \; ,
\label{eq:b7a}
\en
which is also diagonal.
Since the matrices $\hat{G}_{k}$ form an SU(2) algebra the
non-zero coefficients $G^{a}_{i}$ can be chosen to form an
orthogonal $3 \times 3$ matrix. Defining $G^{a}_{i}= b_{ia}$
for the  three values for which $G^{a}_{i} \not= 0$ we
express the generators
\be
T^{c} \; = \; (b^{-1})_{ck} \, L_{k} \; = \; b_{kc} \, L_{k}
\; = \; G^{c}_{k} \, L_{k} \; .
\en
For these 3 generators we then find
\be
- \tr (T^{c} M^{-1} \, i \hat{G}_{l} ) \; = \; G^{c}_{k} \,
\tr ( L_{k} M^{-1} L_{l} ) \; .
\en
Decomposing $L_{i}$ into the $\nu $ irreducible representations
and using (\ref{eq:b7a}) gives
\be
- \tr (T^{c} M^{-1} \, i \hat{G}_{l} ) \; = \; G^{c}_{k} \,
\delta _{lk} \; \sum_{\rho =1}^{\nu } (2 L^{(\rho )} +1 )
\; = \; G^{c}_{l} \; const. \; ,
\en
which provides the desired result.
\end{document}